\documentclass[twocolumn, nolongbibliography]{revtex4-2} 

\usepackage{booktabs}
\usepackage{amsmath}
\usepackage{amssymb}
\usepackage{bm}
\usepackage{color}
\usepackage{graphicx}
\usepackage{xspace}

\newcommand{\Tc}{$T_\text{c}$\xspace}

\begin{document}
\title{Data analysis of $ab$ $initio$ effective Hamiltonians in iron-based superconductors \\--- Construction of predictors for superconducting critical temperature}
  \author{Kota Ido$^{1}$, Yuichi Motoyama$^{1}$, Kazuyoshi Yoshimi$^{1}$ and Takahiro Misawa$^{1,2}$}
\affiliation{$^1$Institute for Solid State Physics, University of Tokyo, 5-1-5 Kashiwanoha, Kashiwa, Chiba 277-8581, Japan}
\affiliation{%
$^2$
Beijing Academy of Quantum Information Sciences, Haidian District, Beijing 100193, China
}

\begin{abstract}
  High-temperature superconductivity occurs in strongly correlated materials 
such as copper oxides and iron-based superconductors. 
  Numerous experimental and theoretical works have been done to identify the key parameters that induce high-temperature superconductivity. 
  However, the key parameters governing the high-temperature superconductivity remain still unclear, which hamper the prediction of superconducting critical temperatures ({\Tc}s) of strongly correlated materials. 
  Here by using data-science techniques, we clarified how the microscopic parameters in the $ab$ $initio$ effective Hamiltonians 
  correlate with the experimental {\Tc}s in iron-based superconductors.
We showed that a combination of microscopic parameters can characterize the compound-dependence of \Tc using the principal component analysis.
  We also constructed a linear regression model that reproduces the experimental \Tc from the microscopic parameters.
  Based on the regression model, we showed a way for increasing \Tc by changing the lattice parameters. 
  The developed methodology opens a new field of materials informatics for strongly correlated electron systems. 
\end{abstract}

\maketitle
%

\section{Introduction}

The discovery of the high-\Tc superconductivity in the copper oxides 
has inspired a huge amount of studies to clarify
the relation between electronic correlations and
high-\Tc superconductivity\cite{Keimer2015}. 
In 2008, the discovery of another high-\Tc superconductivity in iron-based superconductors renewed 
the interest in the superconductivity induced by the electronic correlations\cite{Kamihara2008}.
Theoretical and experimental studies
of these high-\Tc superconductors 
have revealed that 
electronic correlations such as Coulomb interactions are key factors that stabilize
the high-\Tc superconductivity.\cite{RevModPhys.83.1589,HOSONO2015399,si2016high}
However, it remains unclear how the electronic correlations correlate with \Tc.

Recent theoretical progress enables us 
to determine the microscopic parameters in the low-energy effective Hamiltonians 
of solids in an $ab$ $initio$ way. 
The method is often referred to as the $ab$ $initio$ downfolding method\cite{Imada2010}.
Through this method, based on the band structures obtained from
the density functional theory (DFT) calculations, 
screened interactions (e.g., Coulomb interactions)
in the low-energy effective Hamiltonians are calculated 
using the constrained random phase approximation (cRPA) method\cite{Aryasetiawan2004}.
By solving the low-energy effective Hamiltonians,
we can evaluate the physical properties of strongly correlated materials 
in a fully $ab$ $initio$ way beyond the DFT calculations.
It has been shown that the scheme can reproduce
the ground-state phase diagrams of the high-\Tc superconductors\cite{Misawa2014,Ohgoe2020}.

The combination of an
$ab$ $initio$ downfolding scheme and an accurate 
analysis of the derived Hamiltonians 
is useful for clarifying the mechanism of the high-\Tc superconductivity.
However, application of this method has been limited to a few compounds due to
the high numerical cost associated with the accurate analysis of the effective Hamiltonians.
The difficulty of solving the effective Hamiltonians prevents us from gaining a
deep understanding of the high-\Tc superconductivity 
and designing new high-\Tc superconductors.
Therefore, 
to predict new superconductors based on the theoretical calculations,
development of an alternative and efficient method without solving $ab$ $initio$ Hamiltonians is desirable.

Data-driven approaches, such as high-throughput screening, regression analysis, and multivariate statistics, have been used for discovering and designing new compounds with high functionality\cite{Himanen2019}.
Stimulated by the recent progress of data-science,
data-driven analyses focused on superconductors have been 
performed\cite{Klintenberg2013possible,Isayev2015materials,Owolabi2015estimation,Stanev2018machine,Matsumoto2019acceleration,Konno2021deep,Roter2020,Liu2020,Akomolafe2021}.
For example, in Ref. \cite{Stanev2018machine}, utilizing the existing database,
the authors constructed regression models for predicting \Tc based 
on the information of the chemical-composition data of the compounds and the DFT calculations.
Their analysis offers several important insights into
the relation between the chemical composition and \Tc.
However, the effects of electronic correlations in the compounds
have not been directly considered in the existing database.
Electronic correlations are key factors that stabilize the high-\Tc superconductivity
and hence 
the analysis of datasets that explicitly include the 
information about the electronic correlations is desirable.

From the early stage of studies of the iron-based superconductors,
it has been demonstrated that the $ab$ $initio$ approach such as the derivation of the
band structures and the effective Hamiltonians 
is useful for clarifying the magnetism~\cite{Singh_PRL2008,ZPYin_PRL2008} and superconductivity~\cite{Kuroki_RPL2008,Mazin_PRL2008, Kuroki_PRB2009}.
In particular, $ab$ $initio$ evaluations of the microscopic parameters
in the effective Hamiltonians give an insight in understanding the
variety of the iron-based superconductors~\cite{miyake2010comparison}.
In this paper, 
to elucidate the relation between the compound dependence of \Tc
and the microscopic parameters, 
we derive the low-energy effective Hamiltonians 
for 32 iron-based superconductors and 4 related compounds. 
These superconductors offer an ideal platform 
for examining the relation between \Tc and the correlation effects
because \Tc and the strength of the correlations largely depend on the compounds.

Based on the obtained low-energy effective Hamiltonians,
rather than directly solving the Hamiltonians,
we analyze the relations between the experimentally observed
\Tc and the microscopic parameters comprising these Hamiltonians.
This analysis is performed using the data-science techniques
such as principal component analysis (PCA) and 
construction of a linear regression model.
Through the PCA, we find that the 
compound dependence is well characterized by the
first and second principal components
and that the first (second) component consists mainly of the Coulomb interactions (hopping parameters).
For example, we find that 1111 compounds have a similar 
first component and the difference in these compounds
is characterized by the second component. 
We also show that the compound dependence of \Tc 
with respect to the second principal component is manifested as a dome structure.
This result indicates that the microscopic parameters comprising the effective Hamiltonians
provide sufficient information for describing the compound dependence.

We also construct a linear regression model, which
reproduces the experimentally observed \Tc
from the microscopic parameters comprising the effective Hamiltonians.
The obtained model succeeds in reproducing the 
materials dependence of \Tc, as evidenced by a high coefficient of determination ($R^2\sim0.92$).
We show that our regression model can predict a way to enhance
the \Tc of LaFeAsO by changing the structure of the material. 

This paper is organized as follows:
In Sec.~\ref{sec_method}, we denote the basics of the target materials
and explain the $ab$ $initio$ downfolding method.
In Sec.~\ref{sec_derive}, we show the compound dependence of the microscopic
parameters including the electronic correlations in the obtained low-energy effective Hamiltonians. 
In Sec.~\ref{sec_data}, we perform PCA aimed at identifying the main parameters that describe 
the iron-based superconductors using the obtained Hamiltonians.
We also construct a regression model
that reproduces \Tc from the microscopic parameters.
Based on this model, we show a way to optimize \Tc of the iron-based superconductors.
A summary and issues that will be addressed in future work are presented in Sec.~\ref{sec_summary}.
\section{Method} \label{sec_method}
In this section, we briefly explain the numerical methods we employed, 
namely the $ab$ $initio$ downfolding method, the PCA, 
and the construction of a linear regression model.

\subsection{$Ab$ $initio$ downfolding method}
\label{sec_downfolding}
The DFT calculations of the target materials are performed using QUANTUM ESPRESSO\cite{Giannozzi2017}.
As the pseudopotential set, we use sg15 library\cite{Schlipf2015}, which includes the optimized norm-conserving Vanderbilt pseudopotentials (ONCVPSP)\cite{Hamann2013}.
We replace the pseudopotentials for Gd, Nd, Pr, Ce, and Sm, and Tb with that of La to 
eliminate the $f$ metallic bands around the Fermi energy.
This approach has been already applied to derivation of $ab$ $initio$ parameters for cuprates with a lanthanoid\cite{Jang_srep2016}.
We use a $12 \times 12 \times 12$ $k$-mesh in the first Brillouin zone to perform self-consistent calculations. 
The optimized tetrahedron method is employed for the DFT calculations\cite{Kawamura2014}.
An energy cutoff of 120 (480) Ry is set for plane waves (charge densities).

To derive effective Hamiltonians for our target materials, 
we use RESPACK~\cite{Nakamura2021}.
In this work, we construct effective models that are the 
ten-orbital models consisting of 3$d$ orbitals associated with transition metals (TMs), e.g., Fe, Mn, and Ni. 
The Hamiltonians are given as follows.
\begin{align}
&H=H_{\rm 1body}+H_{\rm 2body}, \\
&H_{\rm 1body}=\sum_{i,j}t_{ij,\sigma}c^{\dagger}_{i\sigma}c_{j\sigma}, \\
&H_{\rm 2body}=\frac{1}{2}\sum_{i,j,\sigma,\rho} \Big(
V_{ij}c^\dagger_{i\sigma}c^{\dagger}_{j\rho}c_{j\rho}c_{i\sigma} \notag \\
&+J_{ij}
(c^\dagger_{i\sigma}c^{\dagger}_{j\rho}c_{i\rho}c_{j\sigma}
+c^\dagger_{i\sigma}c^{\dagger}_{i\rho}c_{j\rho}c_{j\sigma})\Big),
\end{align}
where $c_{i\sigma}^{\dagger}$ ($c_{i\sigma}$ ) denotes
the creation (annihilation) operator of an electron with spin $\sigma$ at the $i$-th maximally localized Wannier function (MLWF)\cite{Marzari1997,Souza2001}.
The indices of the MLWFs include the position of the unit cell $\bm{R}$, the orbital degrees of freedom $\mu$ and the index of the TM $n$, i.e., $i=(\bm{R},\mu, n)$. 
Note that the unit cell contains two TMs.
$t_{ij,\sigma}$ in the one-body part of the effective Hamiltonians ($H_{\rm 1body}$) represents the hopping parameters of an electron with spin $\sigma$ between the $i$-th and $j$-th MLWFs.
When we construct the MLWFs, the $x$-$y$ axis is rotated by $45^{\circ}$ from the $a$-$b$ axis of the unit cell.
For this condition, the $x$-$y$ axis is parallel to the nearest TM-TM directions.
Two-body interactions such as Coulomb interactions ($V_{ij}$)
and the exchange interactions ($J_{ij}$) are obtained via cRPA calculations\cite{Aryasetiawan2004}.
We set an energy cutoff of 20 Ry for the dielectric function and calculate dielectric functions using more than 100 bands.
In both calculations, we use  $6 \times 6 \times 6$ $k$ points 
for the 122 systems and $8 \times 8 \times 4$ $k$ points for the others.

The $ab$ $initio$ downfolding method based on cRPA 
has been applied to a wide range of the correlated 
electron systems including iron-based superconductors~\cite{miyake2010comparison,Aichhorn_PRB2010} 
and cuprates~\cite{Jang_srep2016,Hirayama_PRB2019,Nilsson_PRB2019}.
Through these applications, it is shown that the downfolding method 
gives realistic values of the Coulomb and exchange interactions.
However, there are several variations 
of the cRPA method such
as the treatment of the band disentangle~\cite{Souza2001,Miyake_PRB2009,Sasioglu2011}. The quantitative results may depend on the details of the methods.
For example, in the derivation of the single-band Hubbard-type Hamiltonians of the cuprates, 
it is pointed out~\cite{Nilsson_PRB2019} that the values 
of onsite Coulomb interactions depend on the downfolding methods~\cite{Jang_srep2016,Hirayama_PRB2019,Nilsson_PRB2019}.
We expect that this subtle problem does not exist in the derivation of the
ten-orbital Hamiltonians in the iron-based superconductors
since the ten bands cover almost all the degrees of freedom near the Fermi energy.
Therefore, we use the cRPA method with the conventional 
band-disentanglement treatment~\cite{Souza2001} implemented in RESPACK
to capture the overall trend of the compound dependence of the effective Hamiltonians.

For the 122 family, an $ab$ $initio$ model including $d$ orbitals of alkaline earth metals such as 
Ba is more appropriate (than other models) because these bands are located above but near the Fermi level\cite{Usui2019}. 
Nevertheless, using 10 $d$ orbitals in TM atoms, we can construct the MLWFs that reproduce the DFT 
band structure below the energy associated with the $d$ band of alkaline earth metals.

Using the derived microscopic parameters, we employ the following typical parameters as descriptors for the data analysis, 
\begin{align}
  &t_\text{p,q} = g_\text{p} \left( |t_{(0,\mu,0),(0,\nu,1)}| \cdot \Delta_\text{q}(\mu,\nu)  \right), \\
  &t'_\text{p,q} = g_\text{p} \left( |t_{(0,\mu,0),(\bm{R}_a,\nu, 0)}| \cdot \Delta_\text{q}(\mu,\nu)  \right), \\
  &U_\text{p,q} = g_\text{p} \left( V_{(0,\mu,0),(0,\nu,0)} \cdot \Delta_\text{q}(\mu,\nu) \right),  \\
  &V_\text{p,q} = g_\text{p} \left( V_{(0,\mu,0),(0,\nu,1)} \cdot \Delta_\text{q}(\mu,\nu) \right), \\
  &V'_\text{p,q} = g_\text{p} \left( V_{(0,\mu,0),(\bm{R}_a,\nu,0)} \cdot \Delta_\text{q}(\mu,\nu) \right),  \\
  &J_\text{p,q} = g_\text{p} \left( J_{(0,\mu,0),(0,\nu,1)} \cdot \Delta_\text{q}(\mu,\nu)  \right),
\end{align}
where $\bm{R}_a$ is a primitive translation vector along the $a$ axis.  
  $\Delta_\text{q}(\mu,\nu)$ specifies orbital information; $\Delta_\text{all}(\mu,\nu) = 1, \Delta_\text{diag}(\mu,\nu) = \delta_{\mu,\nu}$, and $\Delta_\text{offdiag}(\mu,\nu) = 1-\delta_{\mu, \nu}$.
$g_\text{p}$ is a function for extracting a feature of the derived microscopic parameters among all combinations of $\mu$ and $\nu$.
We use three types of $g_\text{p}$: a function that returns the maximum value (p=max), a function that returns the mean value (p=mean), and a function that returns the minimum value (p=min).
We exclude $t_\text{min}$ and $t'_\text{min}$ because these are zero. 
Throughout this paper, $t_\text{max}=t_\text{max, all}$, $\bar{U}=U_\text{mean, diag}$, $\bar{V}=V_\text{mean, diag}$, and $\bar{J}=J_\text{mean, offdiag}$ for simplicity.

\subsection{PCA}
It is expected that the difference in the parameters of the derived effective Hamiltonians 
can characterize various target materials.
To verify this hypothesis, we visualize the relationship between the parameters and 
extract the main parameters that govern the materials dependence via PCA. 
The PCA, which has been applied to a dataset obtained for superconductors~\cite{Rajan2009}, 
leads to a reduction in the dimensionality of a dataset, thereby providing 
information on the principal components of the target dataset.
We note that the results of 
PCA depend on the employed data set. In this study, we perform PCA analysis
for the Hamiltonians of the compounds with high \Tc to clarify the
microscopic parameter dependence of \Tc. 

Here, we briefly explain {basics of the PCA~\cite{Bishop2006}}.
Using the descriptor vector $\boldsymbol{x}^{(n)}$ of the $n$-th data 
(where $n$ corresponds to {the index} of materials),
we construct the covariance matrix $C$, which is defined as
\begin{align}
  C_{ij} = \frac{1}{N-1}\sum_{n}^{N} x_{i}^{(n)}x_{j}^{(n)},
\end{align}
where {$N$} is the {total} number of the materials.
We note that the average value of the descriptors 
over the materials is set to 0 ($\sum_{n}x_{i}^{(n)}=0$) and
its standard deviation $\sqrt{\sum_{n}\left(x_{i}^{(n)}\right)^2/N}$ is set to 1.
By diagonalizing $C$, 
we obtain the principal vectors $\boldsymbol{v}^{m}$
with the principal values $\lambda_{m}$ ($\lambda_{m}$'s are sorted in descending order).
The first principal vector $\boldsymbol{v}^{1}$
represents the direction associated with the most diverse data.
Using the principal vectors $\boldsymbol{v}^{m}$,
we can project the descriptor vectors onto 
the principal vectors and obtain the $m$th principal value $y_{m}^{(n)}$ as follows,
\begin{align}
y_{m}^{(n)}=\sum_{i} x_{i}^{(n)}v_{i}^{m}.
\label{eq:PCA}
\end{align}
In Sec. \ref{sec_data}, we show that 
the materials dependence can be characterized by the principal values.

\subsection{Construction of regression model}
We construct a linear regression model $f$ for reproducing \Tc
from the parameters of the effective Hamiltonian,
\begin{equation}
  f(\boldsymbol{x}; \boldsymbol{w}) = w_0 + \sum_{i=1}^{N_\text{f}} x_i w_i.
\end{equation}
To introduce non-linearity, 
we inserted the quadratic terms and the ratio terms (such as $t_\text{max}\bar{U}$ and $\bar{V}/U_\text{max}$) 
between the original parameters. 
(please see Appendix \ref{App} for further details of the descriptors).

\begin{figure}[bt!]
  \begin{center}
    \includegraphics[width=8cm]{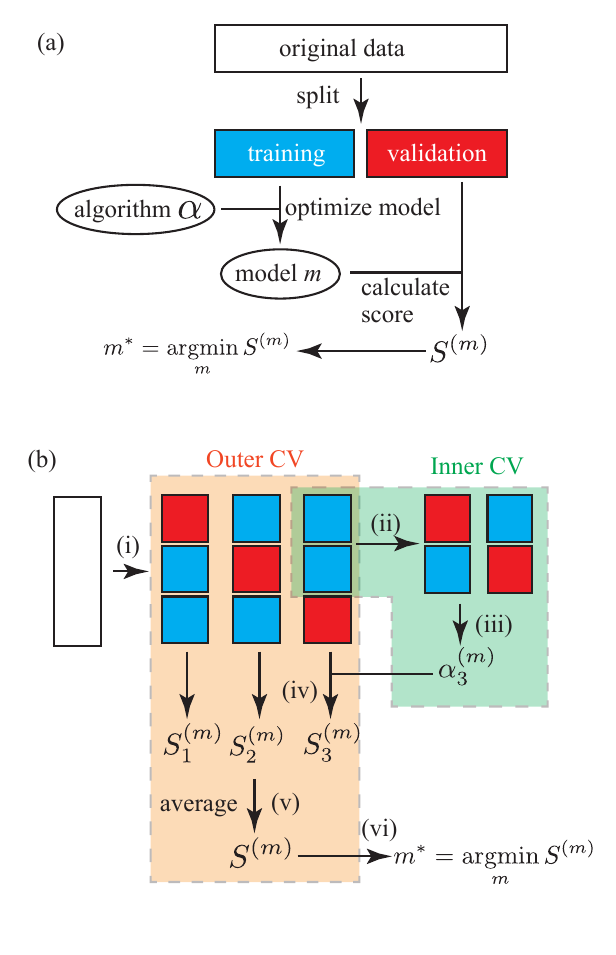}
  \end{center}
  \caption{
    (color online)
    Schematic picture of the validation methods.
    (a) Hold-out validation, which is a basic method for validating. In this method, we split the original data (white rectangle) into the training data (blue) and the validation data (red). 
    Each model $m$ is optimized by the training dataset and the training algorithm with the hyperparameter $\alpha^{(m)}$.
    Once the model is optimized, the score of the model $S^{(m)}$ is calculated by using the validation data.
    (b) Nested CV, which is the method used in the present study. Details of each step (i)-(vi) are
    explained in main text.
}
\label{fig:CV}
\end{figure}

The number of samples in our dataset is comparable to that of the descriptors,
and hence we should carefully control the number of the descriptors in the 
regression models to avoid overfitting.
Therefore, we prepared models where some descriptors were dropped 
in a systematic way --- for example, one model has only hopping parameters as descriptors, 
another one has hopping and onsite-Coulomb interactions, and yet another one has hopping, 
onsite-Coulomb and the cross term between 
these parameters (for details, see Appendix~\ref{App}).
To identify the optimal model among these models, we calculate the \textit{score}s associated with each model. 
One of the simplest ways to estimate score is the hold-out validation (HV) method [Fig.~\ref{fig:CV} (a)]. 
For this method, we first split the original dataset (white boxes in Fig.~\ref{fig:CV}) into two parts, the training dataset (blue boxes) and the validation dataset (red boxes).
The model is optimized by using the training dataset and a training algorithm with the hyperparameter $\alpha^{(m)}$.
The score of the model $m$ is calculated as follows, 
\begin{align}
S^{(m)}_{\rm HV} = \sum_{n\in D_\text{validation}}
\frac{\left[T_\text{c}^{(n)} - f(\boldsymbol{x}^{(n)};\boldsymbol{w}^{(m)})\right]^2}{N_\text{validation}},
\end{align}
where $D_\text{validation}$ is the 
validation dataset, $T_\text{c}^{(n)}$ is the experimental \Tc of the material labeled with $n$ and $N_\text{validation}$ 
is the number of the validation data. 
The model with a minimum score is chosen as the optimal model.

While the HV method has a bias from 
how to split the dataset, the cross-validation (CV) method reduces it.
In this study, we employ the nested CV method~\cite{stone1974cross}, whose procedure is shown in Fig.~\ref{fig:CV} (b). 
In the following, we explain the procedure of the nested CV method.

(i)~{\bf Outer CV}: We first split the original dataset by the leave-one-out CV (LOOCV) method, 
which corresponds to the outer CV. In the outer CV, each model 
 $(m)$ is trained by an algorithm with its hyperparameter $\alpha^{(m)}$.

(ii)~{\bf Inner CV}: To tune $\alpha^{(m)}$ for $k$-th split dataset as $\alpha_k^{(m)}$,
we again apply the LOOCV method to the training data with $N-1$ samples of $
k$-th training dataset, 
which is called the inner CV.

(iii)~{\bf Determination of $\alpha$}:
In this inner CV process, we optimize the weights of the model $\boldsymbol{w}^{(m)}_{kl}$ by minimizing LASSO (least absolute shrinkage and selection operator)\cite{tibshirani1996regression} type cost function;
\begin{align}
&L(\boldsymbol{w}^{(m)}_{kl}; \alpha^{(m)}_k) = \nonumber \\
& \sum_{n\in D_\text{training}^{kl}}
\frac{\left[T_\text{c}^{(n)} - f(\boldsymbol{x}^{(n)};\boldsymbol{w}^{(m)}_{kl})\right]^2}{N_\text{training}^{l}} + \alpha^{(m)}_k |\boldsymbol{w}^{(m)}_{kl}|,
\end{align}
where $D_\text{training}^{kl}$ is the $l$-th training dataset in the inner CV and $N_\text{training}^{l}$ is the number of data in $D_\text{training}^{kl}$,
and the coefficient of the $L_1$ term, $\alpha^{(m)}_k$ is a hyper parameter of LASSO,
which will be tuned by the inner CV.
By using obtained $\boldsymbol{w}^{(m)}_{kl}(\alpha^{(m)}_k)$, the score for the model $m$ and $l$-th dataset split from $k$-th dataset in the outer CV,
$\sigma^{(m)}_{kl}(\alpha^{(m)}_k)$ is calculated as
\begin{equation}
\sigma^{(m)}_{kl}(\alpha^{(m)}_k) =  \sum_{n\in D_\text{validation}^{kl}}
\frac{\left[T_\text{c}^{(n)} - f\left(\boldsymbol{x}^{(n)};\boldsymbol{w}^{(m)}_{kl}(\alpha^{(m)}_k)\right)\right]^2}{N_\text{validation}^{l}}.
\end{equation}
After scores $\sigma^{(m)}_{kl}(\alpha^{(m)}_k)$ for all inner CV dataset are obtained,
the total score of $\alpha^{(m)}_k$
\begin{equation}
  \sigma^{(m)}_{{\rm CV},k}(\alpha^{(m)}_k) = \frac{1}{N-1}\sum_{l=1}^{N-1} \sigma^{(m)}_{kl}(\alpha^{(m)}_k)
\end{equation}
is calculated.
Finally, the hyperparameter $\alpha^{(m)}_k$ is tuned to minimize $\sigma^{(m)}_{{\rm CV},k}(\alpha^{(m)}_k)$.
Note that the $k$-th dataset split by the outer CV is split further into $N-2$ 
samples for training ($N_\text{training}^{l}=N-2$) and one sample for validation ($N_\text{validation}^{l}=1$) in the inner CV.
We use scikit-learn~\cite{scikit-learn} package to solve the LASSO and
use Optuna~\cite{optuna_2019} to optimize $\alpha^{(m)}_k$ 
by means of the Bayesian optimization method.

(iv)~{\bf Evaluation of score}: After $\alpha^{(m)}_k$ is optimized, 
the weight of the model $m$, $w_k^{(m)}$ in 
following equation is again optimized
by LASSO with $\alpha^{(m)}_k$ as in the normal CV method, 
and then the validation score $S_k^{(m)}$ defined in Eq. (\ref{val_score}) is evaluated as follows:
\begin{align}
&S^{(m)}_{k} = \sum_{n\in D_\text{validation}^k} \frac{\left[T_\text{c}^{(n)} - f(\boldsymbol{x}^{(n)};\boldsymbol{w}^{(m)}_k)\right]^2}{N_\text{validation}}, \label{val_score}
\end{align}
where $D_\text{validation}^k$ is the $k$-th validation dataset split by the outer CV and $N_{\text{validation}} = 1$.

(v)~{\bf Average of score}: The final score $S^{(m)}_{\rm CV}$ is then calculated 
from the following equation.
\begin{align}
&S^{(m)}_{\rm CV} = \frac{1}{N} \sum_{k=1}^{N} S^{(m)}_k. \label{final_score}
\end{align}
(vi)~{\bf Determination of the best model}:
By performing the nested CV for all the models we define, 
we obtain the best model $m^*$ with the lowest 
score, i.e. $m^*=\text{argmin}_m S^{(m)}_{\rm CV}$. 

\section{Results} 
\subsection{$Ab$ $initio$ results} \label{sec_derive} 

We obtain the low-energy effective Hamiltonians 
for the 
11-family (e.g., FeSe), 1111-family (e.g., LaFeAsO), 111-family (e.g., LiFeAs), 122-family (e.g., BaFe$_2$As$_2$), 
and 42622-family (e.g., Ca$_4$Al$_2$O$_6$Fe$_2$As$_2$).
For comparison with the iron-based superconductors, 
we have also obtained the low-energy effective Hamiltonians 
for Mn (e.g., BaMnAsF) and Ni (e.g., LaNiAsO) 
analogs of the iron-based superconductors.
Each target material is categorized as one of the above-mentioned families based on the chemical composition of the materials.

The typical microscopic parameters (such as the maximum values
of the hopping parameters and the averaged onsite-Coulomb interactions)
comprising the low-energy effective Hamiltonians and the families of our target compounds are listed in Table~\ref{parameters_table}.
The numerical conditions of the $ab$ $initio$ calculations are detailed in Sec.~\ref{sec_downfolding}.
The crystal structures of all the target materials 
except for ${\rm Ca_4Al_2O_6Fe_2P_2}$, GdFePO and NdFePO are obtained from the experiments.
Most of them are obtained from the Inorganic Crystal Structure Database (ICSD)\cite{bergerhoff1983inorganic,zagorac2019recent} and SpringerMaterials\cite{springer_materials}.
We note that we employ a tetragonal crystal phase at ambient pressure.
For calculations of ${\rm Ca_4Al_2O_6Fe_2P_2}$ (GdFePO and NdFePO), the fractional coordinates of 
the atoms are used as the same as those of ${\rm Ca_4Al_2O_6Fe_2As_2}$(PrFePO), and the lattice constant is 
employed from the experiment.

We find that the amplitudes of the hopping parameters and
the Coulomb interactions exhibit a strong dependence on the compounds.
For example, 
$\overline{U}/t_{\rm max}$ of FeSe ($\overline{U}/t_{\rm max}$ = 10.11) 
is approximately twice that of GdFePO ($\overline{U}/t_{\rm max}$ = 5.15).
This significant variation indicates that iron-based superconductors 
offer an ideal platform for examining the relation between \Tc and the microscopic parameters.
We note that the overall tendency of the obtained parameters is
consistent with the tendency revealed by previous calculations~\cite{miyake2010comparison, miyake2010_42622}.
In particular, the averaged Coulomb interaction $\overline{U}$ values
 agree (to within 10\%) with those obtained in the previous studies,
 except for BaFe$_2$As$_2$.
The $\overline{U}$ value of BaFe$_2$As$_2$ is $\sim$ 20\% smaller 
than that reported in the previous study~\cite{miyake2010comparison}.
This deviation may have resulted from the difference between 
the disentangle treatment for the Ba $d$ band near the Fermi level. 
We neglect clarification of this difference because (in the present study) we focus on the relation
between the \Tc and overall compound dependence of the microscopic parameters. 

\begin{table*}[t]
  \caption{ 
  Compound-dependence of $ab$ $initio$ parameters for target materials. 
  $t_{\rm max}$ denotes the maximum value of the hopping parameters between the nearest-neighbor transition metals (TMs). 
  $\overline{U}$ and $\overline{J}$ denote the averaged values of the diagonal onsite {Coulomb} interactions and the onsite Hund coupling over the TM $d$ orbitals, respectively. 
  $\overline{V}$ is the averaged value of the diagonal density-density interactions between the nearest-neighbor TM sites. 
  The unit of the interaction and hopping parameters is given in eV.
  The effective interactions $\overline{U}/t_{\rm max}$, $\overline{V}/t_{\rm max}$, and $\overline{J}/t_{\rm max}$ are also listed. Ref shows the reference of the crystal structures we employed.
  Information about the microscopic parameters will be public elsewhere.
}
\begin{tabular}{llcccccccr}
\toprule
  material &       family &  $t_\text{max}$ (eV)&  $\overline{U}$ (eV)&  $\overline{V}$ (eV)&  $\bar{J}$ (eV)& $\overline{U}/t_{\rm max}$~~~& $\overline{V}/t_{\rm max}$~~~& $\overline{J}/t_{\rm max}$~~~& Ref \\
\midrule

FeS              &      11 &      0.46 &                4.53 &            1.20 &                   0.55 &       9.91 &       2.63 &              1.21 &   \cite{str_FeS}\\
FeSe             &      11 &      0.41 &                4.14 &            1.04 &                   0.55 &      10.11 &       2.53 &              1.33 &  \cite{str_FeSe}\\
FeTe             &      11 &      0.37 &                3.15 &            0.73 &                   0.50 &       8.57 &       1.98 &              1.37 & \cite{str_FeTe}\\
\hline 
\hline 
CaFeAsH          &    1111 &      0.37 &                2.71 &            0.70 &                   0.44 &       7.26 &       1.88 &              1.19 & \cite{str_CaFeAsH}\\
SrFeAsF          &    1111 &      0.37 &                3.13 &            0.88 &                   0.47 &       8.44 &       2.38 &              1.26 &  \cite{str_SrFeAsF}\\
CaFeAsF          &    1111 &      0.38 &                3.01 &            0.87 &                   0.46 &       7.96 &       2.30 &              1.21 &  \cite{str_CaFeAsF}\\
\hline 
LaFePO           &    1111 &      0.43 &                2.40 &            0.65 &                   0.40 &       5.59 &       1.51 &              0.92 &  \cite{str_LaFePO}\\
PrFePO           &    1111 &      0.43 &                2.44 &            0.65 &                   0.40 &       5.62 &       1.51 &              0.93 &  \cite{str_GdFePO_SmFePO_PrFePO_NdFePO}\\
NdFePO           &    1111 &      0.44 &                2.44 &            0.66 &                   0.40 &       5.51 &       1.49 &              0.90 &  \cite{str_GdFePO_SmFePO_PrFePO_NdFePO}\\
SmFePO           &    1111 &      0.45 &                2.44 &            0.66 &                   0.40 &       5.45 &       1.47 &              0.89 &  \cite{str_GdFePO_SmFePO_PrFePO_NdFePO}\\
GdFePO           &    1111 &      0.47 &                2.40 &            0.66 &                   0.39 &       5.15 &       1.41 &              0.84 &  \cite{str_GdFePO_SmFePO_PrFePO_NdFePO}\\
\hline 
LaFeAsO          &    1111 &      0.36 &                2.46 &            0.61 &                   0.43 &       6.90 &       1.72 &              1.19 &  \cite{Kamihara2008} \\
CeFeAsO          &    1111 &      0.36 &                2.49 &            0.62 &                   0.43 &       6.88 &       1.72 &              1.18 &  \cite{str_CeFeAsO}\\
PrFeAsO          &    1111 &      0.36 &                2.50 &            0.62 &                   0.43 &       6.91 &       1.72 &              1.19 &  \cite{str_NdFeAsO_PrFeAsO_SmFeAsO_TbFeAsO}\\
NdFeAsO          &    1111 &      0.36 &                2.51 &            0.62 &                   0.44 &       6.95 &       1.71 &              1.22 &  \cite{str_NdFeAsO_PrFeAsO_SmFeAsO_TbFeAsO}\\
SmFeAsO          &    1111 &      0.37 &                2.51 &            0.63 &                   0.43 &       6.84 &       1.71 &              1.17 &  \cite{str_NdFeAsO_PrFeAsO_SmFeAsO_TbFeAsO}\\
GdFeAsO          &    1111 &      0.37 &                2.52 &            0.63 &                   0.44 &       6.73 &       1.67 &              1.16 &  \cite{str_GdFeAsO}\\
TbFeAsO          &    1111 &      0.37 &                2.52 &            0.63 &                   0.43 &       6.76 &       1.69 &              1.15 &  \cite{str_NdFeAsO_PrFeAsO_SmFeAsO_TbFeAsO}\\
\hline 
\hline 
LiFeP            &     111 &      0.50 &                3.09 &            0.90 &                   0.45 &       6.23 &       1.82 &              0.90 &  \cite{str_LiFeP}\\
\hline 
LiFeAs           &     111 &      0.42 &                3.02 &            0.78 &                   0.47 &       7.17 &       1.86 &              1.12 &  \cite{str_LiFeAs} \\
NaFeAs           &     111 &      0.39 &                3.06 &            0.79 &                   0.48 &       7.86 &       2.02 &              1.23 &  \cite{str_NaFeAs}\\
\hline 
\hline 
${\rm KFe_2P_2}$           &     122 &      0.34 &                2.43 &            0.61 &                   0.45 &       7.20 &       1.81 &              1.32 &  \cite{str_KFe2P2, Villars2016:sm_isp_sd_2110357}\\
${\rm SrFe_2P_2}$          &     122 &      0.46 &                2.72 &            0.74 &                   0.43 &       5.86 &       1.59 &              0.93 &  \cite{str_SrFe2P2} \\
${\rm CaFe_2P_2}$          &     122 &      0.42 &                2.81 &            0.60 &                   0.52 &       6.75 &       1.44 &              1.24 &  \cite{str_SrFe2As2} \\
${\rm BaFe_2P_2}$          &     122 &      0.45 &                2.51 &            0.64 &                   0.43 &       5.56 &       1.42 &              0.95 &  \cite{str_BaFe2P2}\\
\hline 
${\rm KFe_2As_2}$          &     122 &      0.47 &                3.07 &            0.94 &                   0.46 &       6.59 &       2.03 &              0.98 &  \cite{str_KFe2As2} \\
${\rm SrFe_2As_2}$         &     122 &      0.37 &                2.62 &            0.64 &                   0.44 &       7.09 &       1.74 &              1.20 &  \cite{str_SrFe2As2} \\
${\rm CaFe_2As_2}$         &     122 &      0.37 &                2.38 &            0.52 &                   0.44 &       6.36 &       1.39 &              1.18 &  \cite{str_CaFe2As2}\\
${\rm BaFe_2As_2}$         &     122 &      0.36 &                2.40 &            0.55 &                   0.45 &       6.59 &       1.50 &              1.22 &  \cite{str_BaFe2As2}\\
\hline 
\hline 
${\rm Sr_4Sc_2O_6Fe_2P_2}$    &   42622 &      0.41 &                2.99 &            0.86 &                   0.45 &       7.31 &       2.11 &              1.09 &  \cite{str_Sr4Sc2O6Fe2P2}\\
${\rm Ca_4Al_2O_6Fe_2P_2}$    &   42622 &      0.42 &                3.00 &            0.93 &                   0.45 &       7.10 &       2.21 &              1.07 &  \cite{str_Ca4Al2O6Fe2P2_Ca4Al2O6Fe2As2}\\
\hline 
${\rm Ca_4Al_2O_6Fe_2As_2}$   &   42622 &      0.41 &                2.95 &            0.90 &                   0.45 &       7.12 &       2.16 &              1.09 &  \cite{str_Ca4Al2O6Fe2P2_Ca4Al2O6Fe2As2}\\
\hline 
\hline 
BaMnAsF          &  Mn &      0.29 &                2.75 &            0.77 &                   0.45 &       9.43 &       2.64 &              1.55 &  \cite{str_BaMnAsF}\\
${\rm BaMn_2As_2}$         &  Mn &      0.31 &                2.33 &            0.57 &                   0.43 &       7.47 &       1.82 &              1.36 &  \cite{str_BaMn2As2}\\
\hline 
${\rm BaNi_2As_2}$         &  Ni &      0.42 &                2.81 &            0.48 &                   0.47 &       6.64 &       1.15 &              1.11 &  \cite{str_BaNi2As2} \\
LaNiAsO          &  Ni &      0.39 &                2.81 &            0.61 &                   0.44 &       7.11 &       1.55 &              1.11 &  \cite{str_LaNiAsO}\\
\bottomrule
\end{tabular}
\label{parameters_table}
\end{table*}

\begin{figure*}[bt!]
  \begin{center}
    \includegraphics[width=17cm,clip]{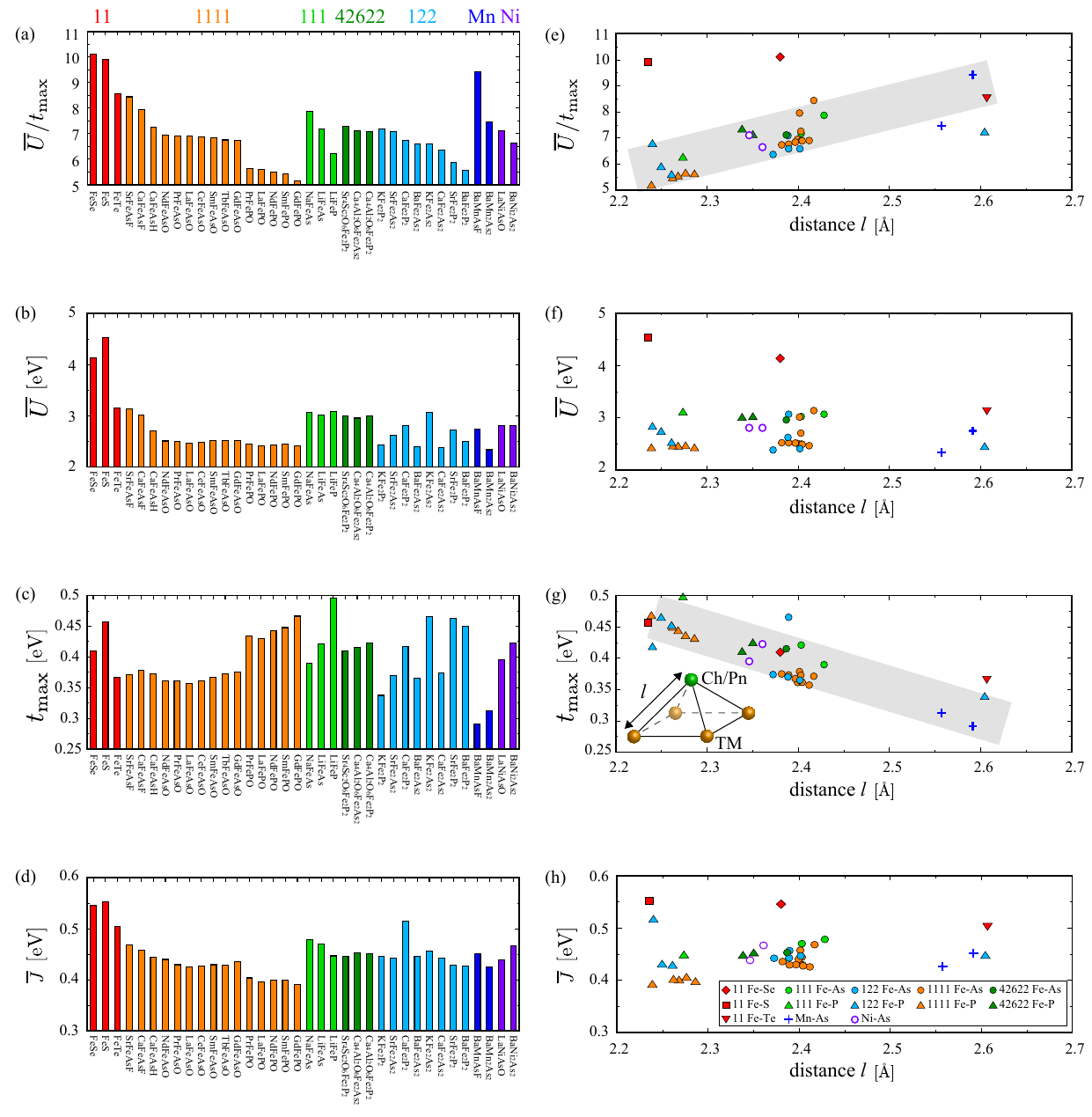}
  \end{center}
  \caption{
  (color online)\textit{Ab initio} parameters for iron-based superconductors and the related compounds. 
  These definitions are explained in the caption of Table \ref{parameters_table}.
  (a-d) Compound-dependence of \textit{ab initio} parameters 
  for (a) $\overline{U}/t_{\rm max}$, (b) $\overline{U}$, (c) $t_{\rm max}$, and (d) $\overline{J}$.
  Colors represent the family name: red bars are the results for the 11 family, orange bars are the results for the 1111 family,
  light-green bars are the results for the 111 family, green bars are the results for the 42622 family, light-blue bars are the results for the 122 family,
  blue bars are the results for the Mn compounds, and purple bars are the results for the Ni compounds.
  (e-h)\textit{Ab initio} parameters as a function of the distance of 
  TM from Ch/Pn. 
  Legends denote the kinds of TM and Ch/Pn of the target materials.
  Filled, open and cross symbols represent the results of Fe, Ni and Mn compounds, respectively.
  For the classification of the family name, the same colors of the symbols are employed as those in the panels (a-d).
  $l$ is the distance between TM and Ch/Pn, which is schematically shown in the inset in panel (g). 
  Gray thick lines in panels (e) and (g) are guides for the eye.
  }
\label{parameters}
\end{figure*}

In Fig.~\ref{parameters} (a-d),
we plot the compound dependence of the typical microscopic parameters.
We find that the effective onsite interaction $\overline{U}/t_{\rm max}$ 
value of the target materials varies significantly (from five to ten).
A significant variation was also noted in Ref.~\cite{miyake2010comparison}
where six iron based superconductors were analyzed. 
The Hund coupling $\overline{J}$ is also dependent on 
the materials, although its range is narrower than that of $\overline{U}$.

From the results obtained for the 1111 family, 
the compounds can be classified into three categories, namely 1111 compounds with P (1111-P), with As (1111-As), and with F or H (1111-F,H).
We find that stronger effective Coulomb interactions $\overline{U}/t_\text{max}$ occur in 1111-F,H than in the other 1111 compounds, owing to
the larger $\overline{U}$ compared with those of 1111-P and 1111-As.
However, the difference between the $\overline{U}/t_\text{max}$ values of 1111-P 
and 1111-As stems from transfer integrals associated with the difference in the localization of MLWF, 
as noted in Ref. \cite{miyake2010comparison}.
A similar tendency is observed for the 111 family.

The $ab$ $initio$ parameters of the relevant compounds are similar to those of iron-based superconductors. 
Mn compounds have strong effective interactions $\overline{U}/t_\text{max}$, 
which arise from the smaller hopping integrals than those of other compounds. 
We note that their occupation numbers in one TM atom are different from those of iron-based superconductors.

Before performing the data-science analyses such as PCA,
we examine the influence of the lattice structure on the
materials dependence of the microscopic parameters. 
The importance of the anion [chalcogen (Ch) or pnictide (Pn)] position is 
pointed out in the previous studies~\cite{Lee2008effect,Mizuguchi2010review}, 
and hence we plot the typical microscopic 
parameters as a function of the distance between Ch/Pn and TM [see Fig.~\ref{parameters}(e-h)].
As shown in Fig.~\ref{parameters} (e), the magnitude of the effective onsite 
Coulomb interaction $\overline{U}/t_{\rm max}$ increases
with increasing distance.
As shown in Fig.~\ref{parameters} (f) and (g),
we find that this trend is mainly governed by $t_{\rm max}$, i.e., 
$t_{\rm max}$ decreases almost monotonically with increasing the distance whereas $\overline{U}$ is almost independent of the distance.
This result indicates that the difference between the distances controls $t_{\rm max}$
mainly through the hybridization between TM $d$ orbitals and Ch/Pn $p$ orbitals.
We note that $\overline{J}$ exhibits no clear distance dependence.
\subsection{Data analysis}
\label{sec_data}
By using the effective Hamiltonians reported in the previous section, 
we perform the PCA and the regression analysis for \Tc of iron-based superconductors.
We focus on the superconductivity of the iron-based superconductors in this study,
since the \Tc values of Ni compounds are very low, and Mn compounds exhibit no superconductivity.
The results obtained from our data analysis are summarized in Table \ref{tc_table}. 
We provide details of the PCA analysis and the construction of the regression model in subsequent sections.

Here, we detail our strategy for assigning \Tc to the effective Hamiltonians.
Since the superconductivity appears in the doped compounds, 
it is the best way to derive the low-energy effective Hamiltonians for the doped compounds.
When we can obtain the information of the lattice structures for both
end compounds (e.g. FeSe and FeTe), we interpolate the microscopic parameters 
of the effective Hamiltonians for the end compounds as shown in Table \ref{tc_table}.
However, information of the lattice structures for the end compounds 
is not available in most cases for 1111 compounds.
Nevertheless, in terms of the microscopic parameters, 
we can expect the difference is small in the low-doping regime where $T_{\text{c}}^{\text{exp.}}$ shown in Table \ref{tc_table} was observed. 
Therefore, we simply assume that the effective Hamiltonian 
of the closest stoichiometry compound 
can govern the superconductivity in the doped compounds. 
As we will show, the PCA analysis shows that 
the combination of the microscopic parameters well describe the compounds dependence of \Tc.
This result supports the validity of the assumption. 
We also note that the same assumption was used in the analysis
of the relation between \Tc and the onsite Coulomb interactions in cuprates~\cite{Jang_srep2016,Nilsson_PRB2019}.

\begin{table*}[t]
  \caption{Superconducting transition temperature $T_c$ values of target materials. $Ab$ $initio$ effective Hamiltonians are used as descriptors for the construction of the regression model. 
  $T_\text{c}^\text{exp.}$ and $T_\text{c}^\text{pred.}$ denote onset \Tc observed in experiments and \Tc predicted by our regression model, respectively.
  $y_1$ and $y_2$ are the values of the 1st and the 2nd principal components in the PCA, respectively.
  $(H_\text{A},H_\text{B})$ denotes the linear interpolation of the parameters between $H_\text{A}$ and $H_\text{B}$, whose ratio corresponds to the stoichiometric ratio of a target material.
 The dataset contains 29 materials.
}
\begin{tabular}{llccccc}
\toprule
material   & family~~& $T_\text{c}^\text{exp.}$ (K)~~~~& $T_\text{c}^\text{pred.}$ (K)               &~~~~~$y_1$   &~~~~~$y_2$      &~~~~~Hamiltonian \\
\midrule
  ${\rm  FeS}$                & 11 & 5~\cite{str_FeS}                     & 6.2  & -14.89 & 2.07   & $H_\textrm{FeS}$  \\ 
  ${\rm  FeSe}$               & 11 & 13.5~\cite{Tc_FeSe_Mizuguchi2008}           & 13.4 & -10.51 & -2.79  & $H_\textrm{FeSe}$ \\ 
  ${\rm  FeSe_{0.4}Te_{0.6}}$ & 11 &23.5\cite{note1,Tc_FeSeTe_Noji2012}  & 24.3 & -4.56  & -4.63  & ($H_\textrm{FeSe}$,$H_\textrm{FeTe}$) \\ 
  \hline
  \hline
  ${\rm Ca_{0.77}La_{0.23}FeAsH}$ & 1111 & 47.4~\cite{Tc_CaFeAsH_Muraba2014} & 55.1     &  0.98 & -1.33       & $H_\textrm{CaFeAsH}$ \\
  ${\rm Sr_{0.5}Sm_{0.5}FeAsF}$   & 1111 & 56~\cite{Tc_SrFeAsF_Wu2009}       & 51.1     & -3.33 & -2.68       & $H_\textrm{SrFeAsF}$  \\
  ${\rm Ca_{0.4}Pr_{0.6}FeAsF}$   & 1111 & 56~\cite{Tc_CaFeAsF_Cheng2009}    & 53.6     & -3.33 & -1.43       & $H_\textrm{CaFeAsF}$  \\
  \hline
  ${\rm LaFePO}$ & 1111 & 6.6~\cite{Tc_LaNdPrFePO_Baumbach2009}              & 3.2      &  3.97 &  2.86       & $H_\textrm{LaFePO}$  \\ 
  ${\rm PrFePO}$ & 1111 & 3.2~\cite{Tc_LaNdPrFePO_Baumbach2009}              & 8.2      &  3.54 &  3.49       & $H_\textrm{PrFePO}$  \\ 
  ${\rm NdFePO}$ & 1111 & 3.1~\cite{Tc_LaNdPrFePO_Baumbach2009}              & 4.6      &  3.46 &  3.65       & $H_\textrm{NdFePO}$  \\ 
  ${\rm SmFePO}$ & 1111 & 3~\cite{Tc_SmFePO_Kamihara2008}                    & 6.4      &  3.39 &  4.13       & $H_\textrm{SmFePO}$  \\ 
  ${\rm GdFePO}$ & 1111 & 6.1~\cite{Tc_GdFePO_Liang2010}                     & 2.5      &  3.33 &  5.90       & $H_\textrm{GdFePO}$  \\ 
  \hline
  ${\rm LaFeAsO_{0.92}H_{0.08}}$   & 1111 & 29 \cite{note2,Tc_LaFeAsO_Iimura2012} & 43.3 & 4.60   & -3.27   & $H_\textrm{LaFeAsO}$ \\ 
  ${\rm CeFeAsO_{0.75}H_{0.25}}$   & 1111 & 47~\cite{Tc_CeFeAsO_Matsuishi2012}            & 46.1 & 4.15   & -2.59   & $H_\textrm{CeFeAsO}$ \\
  ${\rm PrFeAsO_{0.89}F_{0.11}}$   & 1111 & 52~\cite{Tc_PrFeAsO_Ren2008}                  & 48.4 & 4.05   & -2.41   & $H_\textrm{PrFeAsO}$ \\
  ${\rm NdFeAsO_{0.8}F_{0.2}}$     & 1111 & 52.8~\cite{Tc_NdFeAsO_Adamski2017}            & 55.2 & 3.85   & -2.00   & $H_\textrm{NdFeAsO}$ \\
  ${\rm SmFeAsO_{0.917}F_{0.083}}$ & 1111 & 55.6~\cite{Tc_SmFeAsO_Kamihara2010}           & 48.1 & 3.61   & -1.92   & $H_\textrm{SmFeAsO}$ \\
  ${\rm GdFeAsO_{0.85}}$           & 1111 & 53.5~\cite{Tc_GdFeAsO_Yang2008}               & 49.5 & 3.26   & -0.95   & $H_\textrm{GdFeAsO}$ \\
  ${\rm TbFeAsO_{0.9}F_{0.1}}$     & 1111 & 50~\cite{Tc_TbFeAsO_Yates2009}                & 46.9 & 3.23   & -1.49   & $H_\textrm{TbFeAsO}$ \\
  \hline
  \hline
  ${\rm LiFeP}$        & 111  &6~\cite{str_LiFeP}                                 & 5.7  & -5.44  & 6.67    & $H_\textrm{LiFeP}$  \\ 
  \hline
  ${\rm NaFeAs}$       & 111  &26~\cite{Tc_NaFeAs_Wang2009}                               & 24.1 & -1.89  & -3.14   & $H_\textrm{NaFeAs}$ \\ 
  ${\rm Li_{0.6}FeAs}$ & 111  &18~\cite{Tc_LiFeAs_Wang2008}                               & 28.1 & -2.60  & 0.55     & $H_\textrm{LiFeAs}$ \\ 
  \hline
  \hline
  ${\rm  SrFe_2As_{1.3}P_{0.7}}$                   & 122 & 27~\cite{Tc_SrFe2AsP2_Shi2010}               & 31.0 & 2.15  & 0.77   & ($H_\textrm{SrFe$_2$As$_2$}$,$H_\textrm{SrFe$_2$P$_2$}$) \\ 
  ${\rm  BaFe_2As_{1.48}P_{0.52}}$                 & 122 & 31~\cite{Tc_BaFe2AsP2_Kasahara2010}          & 33.0 & 4.73  & -0.64  & ($H_\textrm{BaFe$_2$As$_2$}$,$H_\textrm{ BaFe$_2$P$_2$}$) \\ 
  \hline
  ${\rm  K_{0.5}Sr_{0.5}Fe_2As_2}$                 & 122 & 37~\cite{Tc_KSrFe2As2_Sasmal2008}            & 26.1 & -1.17 & 0.65   & ($H_\textrm{KFe$_2$As$_2$}$,$H_\textrm{SrFe$_2$As$_2$}$)  \\ 
  ${\rm  Ca_{0.83}La_{0.17}Fe_2As_{1.88}P_{0.12}}$ & 122 & 48 \cite{note3,Tc_CaFe2As2_Kudo2013} & 49.0 & 6.17  & -0.79  & $H_\textrm{CaFe$_2$As$_2$}$  \\ 
  ${\rm  Ba_{0.6}K_{0.4}Fe_2As_{2}}$               & 122 & 38.7~\cite{Tc_BaKFe2As2_Rotter2008}          & 29.2 & 1.03  & -0.20  & ($H_\textrm{BaFe$_2$As$_2$}$,$H_\textrm{KFe$_2$As$_2$}$)  \\ 
  \hline
  \hline
  ${\rm Ca_4Al_2O_{5.80}Fe_2P_2}$  & 42622 & 17.1~\cite{str_Ca4Al2O6Fe2P2_Ca4Al2O6Fe2As2}             &  19.2 & -5.29 & 0.66    & $H_\textrm{Ca$_4$Al$_2$O$_6$Fe$_2$P$_2$}$ \\ 
  ${\rm Sr_4Sc_2O_6Fe_2P_2}$       & 42622 & 17~\cite{str_Sr4Sc2O6Fe2P2}                    &  18.8 & -2.11 & 0.98    & $H_\textrm{Sr$_4$Sc$_2$O$_6$Fe$_2$P$_2$}$ \\ 
  \hline
  ${\rm Ca_4Al_2O_{5.65}Fe_2As_2}$ & 42622 & 28.3~\cite{str_Ca4Al2O6Fe2P2_Ca4Al2O6Fe2As2}             &  26.9 & -4.37 & -0.13   & $H_\textrm{Ca$_4$Al$_2$O$_6$Fe$_2$As$_2$}$ \\ 
\bottomrule
\end{tabular}
\label{tc_table}
\end{table*}

\begin{figure}[t!]
  \begin{center}
    \includegraphics[width=8cm,clip]{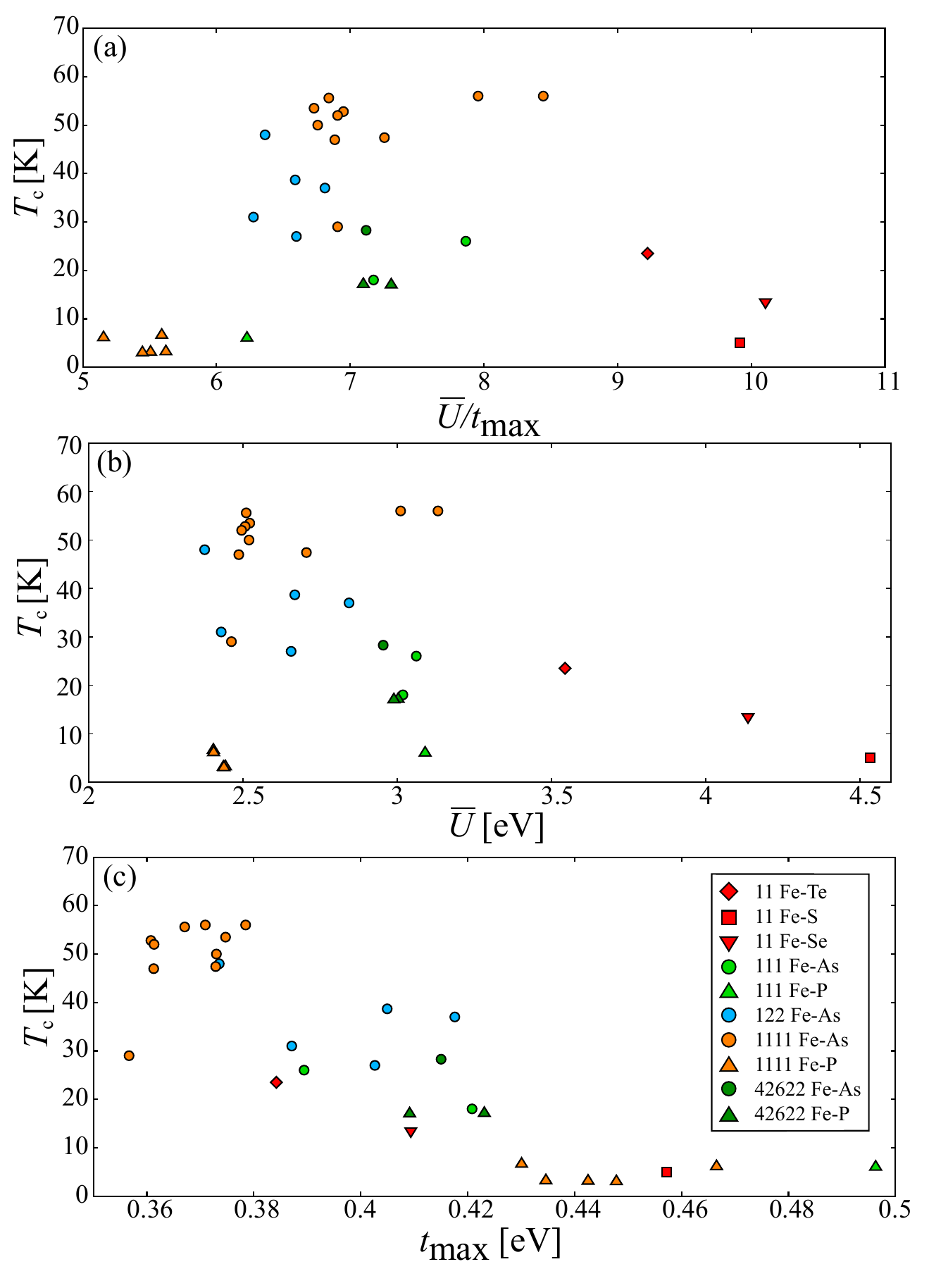}
  \end{center}
  \caption{(color online) Plot of \Tc as a function of (a)$t_{\rm max}$, (b)$\bar{U}$, and (c)$\bar{U}/t_{\rm max}$.}
\label{Tc_Ut}
\end{figure}

\begin{figure}[h!]
  \begin{center}
    \includegraphics[width=8cm,clip]{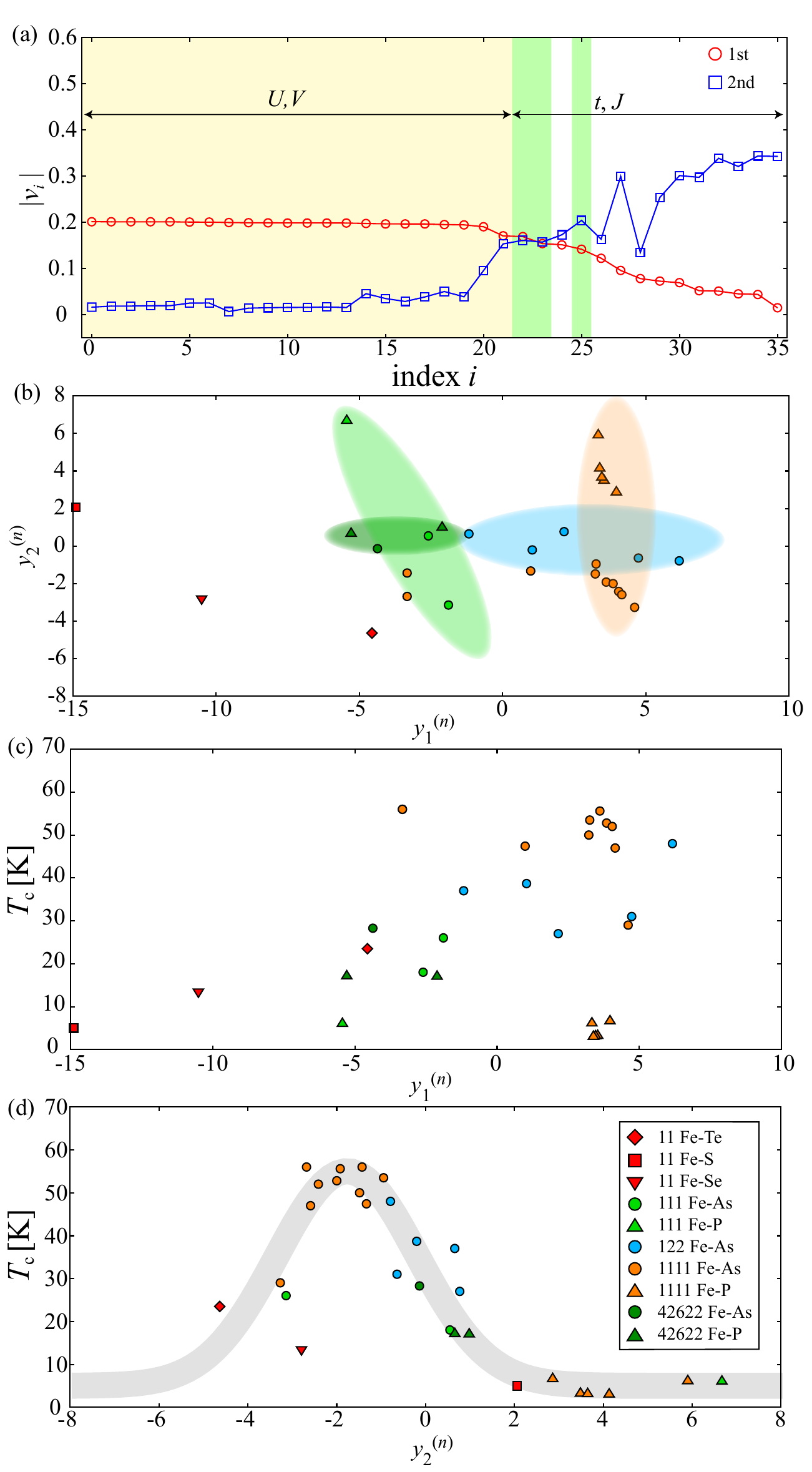}
  \end{center}
\caption{(color online)~
  PCA analysis of the $ab$ $initio$ parameters associated with the iron-based superconductors. 
(a)~Each component of the 1st and the 2nd principal vector.
  The descriptors are sorted in descending order of the absolute value corresponding to the components of the first principal vector, $|v^1_i|$.
The interaction parameters (onsite and offsite Coulomb interactions) govern the 1st principal vector,
while the hopping parameters and Hund couplings govern the 2nd principal vector.
  The descriptors in the yellow, green, and white regions are related to the Coulomb interactions, the Hund coupling, 
and the transfer parameters, respectively. 
  See Appendix~\ref{App} for the features of the descriptors for the PCA and the details of the index $i$.
  (b)~Compound-dependence of the 1st and the 2nd principal components, which are defined in Eq.~(\ref{eq:PCA}).
  Notations are the same as in panels (c) and (d).
  Colored ellipses are guides for the eye.
  (c) and (d)~$T_\text{c}$ as functions of the 1st and the 2nd principal component. 
  Gray thick curves are guides for the eye.
}
\label{pca_all}
\end{figure}

\subsubsection{PCA}
Before showing the results of PCA, we first examine
how \Tc depends on the typical microscopic parameters such as 
$\bar{U}/t_{\rm max}$, $\bar{U}$, and $t_{\rm max}$. 
As shown in Fig.\ref{Tc_Ut},
We do not find any significant correlation between \Tc and $\bar{U}$.
On the other hand, we find that \Tc shows 
a subtle peak structure as a function $\bar{U}/t_{\rm max}$ and $t_{\rm max}$.
This is in sharp contrast with the cuprates where it is proposed that
\Tc has some correlations with both $U$ and $U/t$~\cite{Jang_srep2016,Nilsson_PRB2019}.
As we show later, from the PCA analysis, we show that 
\Tc shows a clear peak structure as a function of the second principal component, 
which mainly consists of the hopping parameters.
Thus, although this $t_{\rm max}$ dependence 
roughly captures some aspects of the compound dependence of \Tc,
the PCA analysis is a more systematic way 
for extracting the important descriptors 
for characterizing the target systems without any prior knowledge.

Here, we explain why the parameter dependence of \Tc is
different between iron-based superconductors and cuprates.
In iron-based superconductors, the hopping parameters, such as $t_{\rm max}$,
mainly change while the Coulomb interactions, such as $U$, 
mainly change in cuprates~\cite{Jang_srep2016,Nilsson_PRB2019, Jean2022,Hirayama2022}.
This difference can be understood as follows: 
In iron-based superconductors, the distance between anions and irons (e.g., As and Fe) widely varies, 
as shown in Fig.~\ref{parameters}.
This indicates that the hybridizations between anions and irons, i.e.,  
the amplitudes of the hopping parameters also widely vary.
In contrast, in cuprates, the distance between Cu and O in the CuO$_{2}$ plane 
does not largely depend on compounds. 
Thus, the amplitude of the hopping parameters does not largely change in cuprates. 
This difference in the lattice structures induces the different parameter dependence of \Tc.   
Nevertheless, we note that the normalized electron correlation $U/t$ 
plays some role in stabilizing high-\Tc superconductivity since $U/t$ changes in both iron-based superconductors and cuprates.

For performing the PCA, 
we employ $36$ microscopic parameters as the descriptors, which consist of the Coulomb interactions ($V_{ij}$),
the Hund couplings ($J_{ij}$), 
and the hopping parameters ($t_{ij}$).
We obtain $\lambda_1=25.06, \lambda_2=7.94, 
\lambda_3=2.12, \lambda_4=1.20$, and  $\lambda_5=0.37$ as the five dominant principal values.
The first and the second principal values
are large compared with the other principal values,
and hence it is expected that the first and the second principal vectors ($\boldsymbol{v}^{1}$ and $\boldsymbol{v}^{2}$) accurately reflect 
the material dependence.
In Fig.~\ref{pca_all}(a),
we plot the absolute value of each component
comprising these vectors.
We find that the interaction parameters (such as onsite and offsite Coulomb
interactions) govern the first principal vector while
the hopping parameters and Hund couplings govern
the second principal vector.

We plot the first and the second principal components ($y_{1}^{(n)}$ and $y_{2}^{(n)}$)
in Fig.~\ref{pca_all}(b).
For 1111 and 111 compounds, 
we find that the second component characterizes
the differences within families, i.e., 
the same family has a similar first component, but
each compound in the family has a different second component.
For example, the first component of the 1111 family is approximately four, but
the second component varies from -3.2 to 6.0.
In contrast,
we find that 122 and 42622 compounds exhibit
opposite tendencies:
the same family has a similar second component, but
each compound has a different first component.
We note that the first component changes significantly 
by changing the total electron density 
for 1111 and 122 families.

The 11 family exhibits exceptional behavior, i.e., 
both the first and the second components 
exhibit considerable compound dependence.
This exceptional behavior may be related to
exotic phenomena such as 
the absence of antiferromagnetic order and the high-$T_\text{c}$
superconductivity in a monolayer occurring in FeSe\cite{hsu2008superconductivity,li2009first,Qing2012}.

In Fig.~\ref{pca_all}(c) and (d),
we plot $T_{\rm c}$ as a function of the first and the second principal components. 
We find that $T_{\rm c}$ is better correlated with the second component than the first component.
In particular, we find that the dependence of $T_{\rm c}$ 
on the second component is manifested as a dome structure reminiscent
of Lee's plot~\cite{Lee2008effect,Mizuguchi2010review}.
This result indicates that the difference in the 
microscopic parameters associated with the effective Hamiltonians
provides sufficient information for explaining the 
compound dependence of \Tc.
By constructing the 
regression model for reproducing \Tc, 
we show that these parameters 
allow quantitative reproduction of this dependence.

\subsubsection{Regression model for estimating $T_\text{c}$}
\begin{figure}[bt!]
  \begin{center}
    \includegraphics[width=8cm,clip]{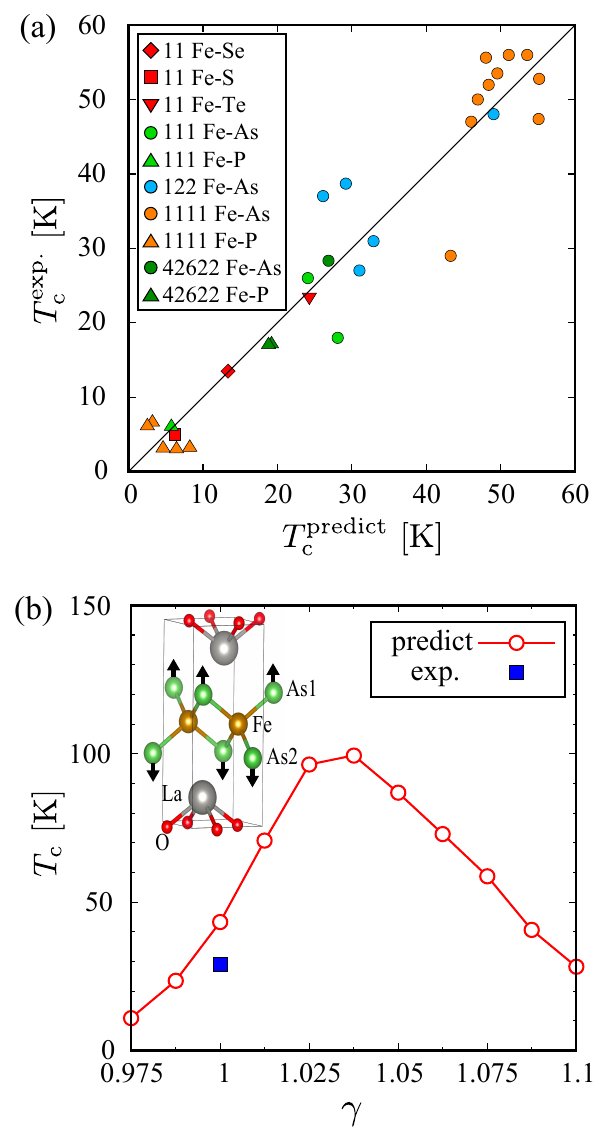}
  \end{center}
  \caption{(color online) (a) Experimental $T_\text{c}$ vs. predicted $T_\text{c}$ obtained from our regression model. 
  Notations are the same as those of Fig. \ref{parameters}.
  The black thin line represents $T_\text{c}^\text{predict} = T_\text{c}^\text{exp}$.
  The coefficient determination $R^2$ is about 0.92.
  Data is also shown in Table~\ref{tc_table}.
  Note that ${\rm  SrFe_2As_{1.3}P_{0.7}}$ and ${\rm  BaFe_2As_{1.48}P_{0.52}}$ are categorized as ${\rm Fe-As}$, and ${\rm  FeSe_{0.4}Te_{0.6}}$ is categorized as ${\rm Fe-Te}$.
  (b) 
  Prediction of $T_\text{c}$ for LaFeAsO with hypothetical structures with different c-axis components of fractional coordinates corresponding to As, $f_{\rm As}$.
  $\gamma$ corresponds to the ratio of $f_{\rm As}$ associated with the original LaFeAsO\cite{Kamihara2008} to that of the hypothetical materials.
  Note that only the material for $\gamma=1$ has been found in the experiments; the corresponding $T_\text{c}$ is plotted as the blue square.
  Inset represents the structure of the original LaFeAsO, which is depicted using VESTA\cite{Momma2011}. 
  The fractional coordinates of two As atoms, As1 and As2, are set to (0.5, 0.0, $\gamma f_{\rm As}$) and (0.0, 0.5, 1-$\gamma f_{\rm As}$), respectively.
  Black thick arrows denote the displacement pattern of the Raman-active ${\rm A_{1g}}$ phonon.
}
\label{reg}
\end{figure}

After determining the best regression model within the nested CV approach explained in Sec.~\ref{sec_method}, 
we obtain our predictor using all the \Tc data obtained in the experiments.
Figure~\ref{reg}(a) shows the accuracy of the obtained regression model.
We find that this model reproduces the experimental \Tc data, as indicated by the high coefficient of determination ($R^2 \approx 0.92$). 
This result confirms that our regression model can predict the $T_\text{c}$ of associated with 
several different families for iron-based superconductors. 

We apply our predictor to low-energy Hamiltonians for a hypothetical compound, 
whose lattice structure is systematically changed from that of LaFeAsO.
We vary the fractional coordinates of two As atoms as (0.5, 0.0, $\gamma f_{\rm As}$) and (0.0, 0.5, 1-$\gamma f_{\rm As}$);
$f_{\rm As}$ is the fractional coordinate of As for the original LaFeAsO, namely $f_{\rm As}=0.6512$ \cite{Kamihara2008}, and the parameter $\gamma$ indicates changes in the height of As from the Fe plane.
A schematic showing this lattice distortion is presented in the inset of Fig.~\ref{reg}(b).
Note that the fractional coordinates of the other atoms and the lattice constants do not change.

We obtain the low-energy effective Hamiltonians for the hypothetical compounds. 
From the microscopic parameters in the Hamiltonians, we calculate the \Tc of each compound using the regression model. 
In Fig.~\ref{reg}(b), we show the $\gamma$ dependence of $T_\text{c}$ 
for hypothetical LaFeAsO materials.
We find that \Tc increases with increasing $\gamma$ and vice versa.
At $\gamma \sim 1.03$, \Tc reaches $\sim$ 100K. 
The applicability range of the regression model is limited to the parameter space around
the existing compounds, and therefore the role of lattice distortion in inducing 
high-\Tc superconductivity ($T_\text{c} \approx 100$K) remains unclear.
Nevertheless, the qualitative behavior of \Tc may be 
correctly predicted by the regression model.
This result indicates that our regression model 
captures the essence of Lee's plot\cite{Lee2008effect,Mizuguchi2010review}
and the theoretical study~\cite{Kuroki_PRB2009}, which
suggests that the height of As from the Fe plane plays a key role in 
determining the compound dependence of \Tc.

\begin{figure}[t!]
  \begin{center}
    \includegraphics[width=8.5cm,clip]{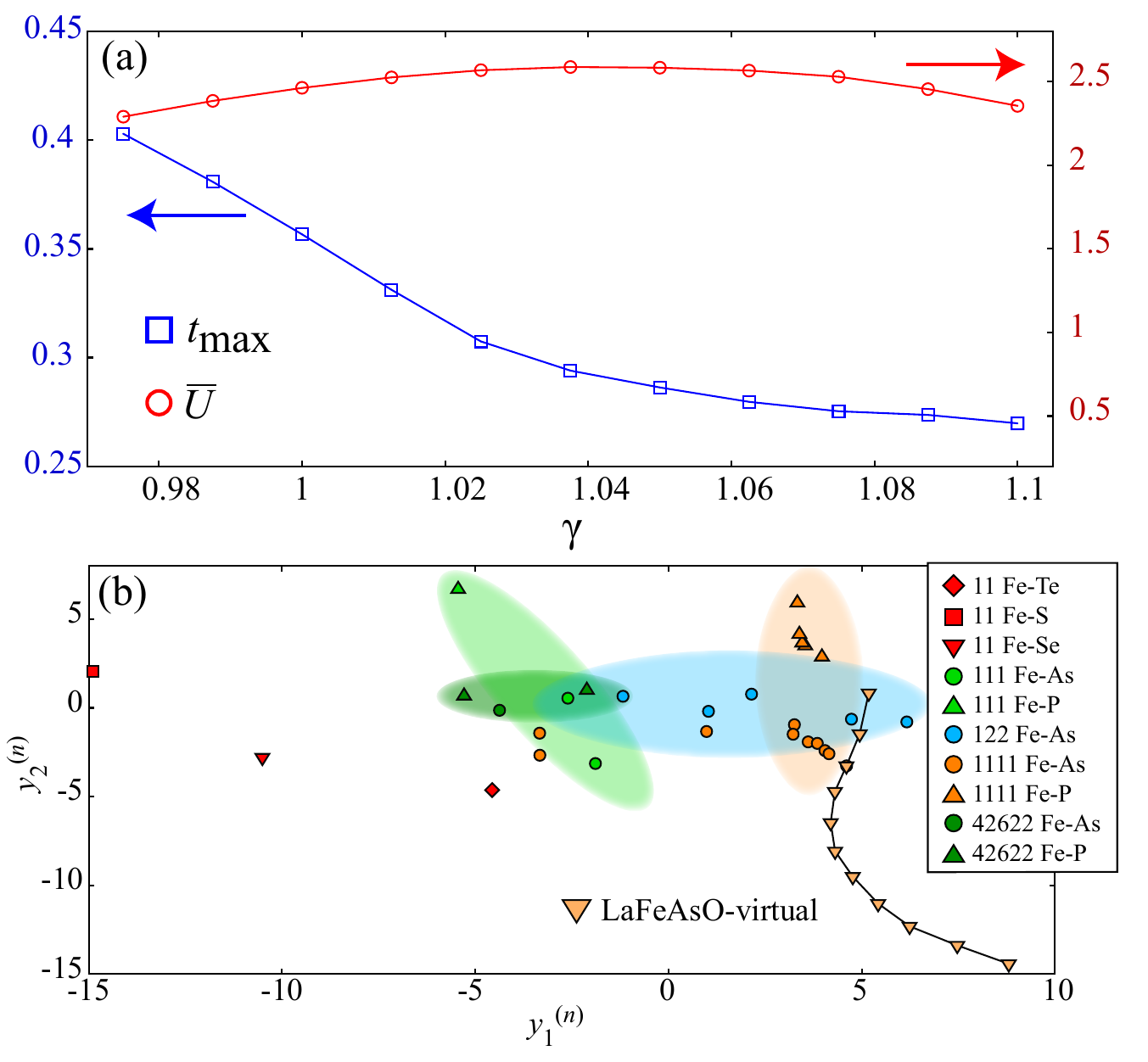}
  \end{center}
  \caption{(color online) (a) $\gamma$ dependence of $\bar{U}$ and $t_{\rm max}$ for the virtual compounds. The units of $\bar{U}$ and $t_{\rm max}$ are eV.
  (b) The first and second principal components for the virtual 
   compounds.~The definition of the principal component is given in Eq.~(\ref{eq:PCA}).~The other points are the same as in Fig. \ref{pca_all}(b). 
}
\label{Virtual_Ut}
\end{figure}

Here, we examine the relation between enhancement of \Tc and the microscopic parameters.
As shown in Fig.~\ref{Virtual_Ut}(a), the maximum value of the nearest hoppings $t_{\rm max}$
decreases by increasing $\gamma$ while $\bar{U}$ has a broad peak around $\gamma\sim1.04$.
This result indicates that we can enhance \Tc of LaFeAsO by increasing
the amplitude of the electronic correlations.
In Fig.~\ref{Virtual_Ut}(b), we plot the first and the second principal components
for the virtual materials. We find that the changes in the second principal component
governs the enhancement of \Tc around LaFeAsO. 
This plot also indicates that LaFeAsO is located at the edge of the existing materials in the $y_{1}-y_{2}$ plane.
Therefore, it is plausible that we can reach the unexplored region and enhance \Tc
by changing the height of As in LaFeAsO.

We also consider the possibility of realizing hypothetical materials.
In our \textit{ab initio} calculations of such materials, we do not perform the structure optimization and simulations for phonon's properties. 
Therefore, our proposed materials may be unattainable for the conventional bulk system at ambient pressure.
Stable materials may be used to obtain an interface structure.
Such interface or thin-film structures induce drastic 
changes in the electronic states of target materials and may enhance \Tc.
Successful experiments applying this approach have already been reported 
for high-\Tc superconductors such as ${\rm La_2CuO_4}$ and FeSe~\cite{Wu2013,Qing2012}.
The material structure can also be controlled via laser irradiation, which has been successfully applied to FeSe\cite{Suzuki2019}.
In that work, strong laser irradiation led to changes in the height of Se atoms from the Fe-Fe plane. 
This results from the displacive excitation of coherent phonon mechanism\cite{Zeiger1992}, which governs the excitation of ${\rm A_{1g}}$ Raman active modes.
These modes also occur in LaFeAsO\cite{Hadjiev2008}, and hence we expect that our predicted structure would be realized via laser irradiation.

\section{Summary}\label{sec_summary}
In summary, we have derived the low-energy effective Hamiltonians 
for 32 iron-based compounds and four related compounds using the $ab$ $initio$ downfolding method.
We have found that microscopic parameters such as the hopping parameters
and the Coulomb interaction are associated with a wide range of iron-based superconductors.
To systematically characterize the compound dependence of the
microscopic parameters, we perform the PCA.
As a result, 
we find that the first principal vector consists of mainly interaction terms, 
while the second consists of mainly hopping terms.
We show that the first principal values characterize the difference within the 42622 and 122 families while
the second principal values characterize the difference within the 1111 and 111 families.
Moreover, we show that the compound dependence of \Tc 
with respect to the second principal value is manifested as a dome structure reminiscent of
Lee's plot~\cite{Lee2008effect,Mizuguchi2010review}.
This result suggests that the obtained microscopic parameters
adequately reflect the compound dependence of \Tc.

Afterward, we construct a regression model for reproducing the compound dependence of \Tc
from the microscopic parameters of the \textit{ab initio} Hamiltonians.
We succeed in reproducing experimental \Tc values of the iron-based superconductors ($R^2\approx 0.92$). 
Using the regression model, we show that 
the \Tc of LaFeAsO can be enhanced by properly changing the As height from the Fe plane.
Such a hypothetical structure would be realized in experiments including laser irradiation processes. 
Applying pressure can also control the anion height from the Fe plane, and this is accompanied by changes in the lattice constants\cite{Okada2008,Okabe2010,kobayashi2016pressure}.
It has been reported that \Tc increases to about 50 K by applying pressure into LaFeAsO\cite{kobayashi2016pressure}. 
However, the current regression model does not have enough accuracy to quantitatively reproduce the experiments because the dataset does not include the experimental data under pressure.
Future studies will determine whether our approach can reproduce \Tc of iron-based superconductors under pressure by updating the dataset.
Using cutting-edge numerical packages (such as mVMC\cite{Misawa2017}) to directly solve the low-energy effective Hamiltonians in order to determine whether \Tc is really enhanced for the hypothetical structures would be an intriguing challenge.
However, the numerical cost of such calculations is rather high and hence this analysis will be considered in future studies.

The present work shows the possibility that the data-science techniques can help us to bypass a difficult problem
--{\it solving the Hamiltonians for quantum many-body systems}-- 
and to clarify the effect of microscopic parameters on superconductivity. 
Further applications of the developed method to other exotic phenomena (such as
the correlated topological materials) 
may be considered in future work.

{\it Acknowledgments.}---
%
We acknowledge Tetsuya Shoji, Noritsugu Sakuma, Mitsuaki Kawamura, Takashi Miyake and Kazuma Nakamura
for fruitful discussions and important suggestions.
We also thank Mitsuaki Kawamura for providing us his scripts to 
generate inputs for Quantum ESPRESSO and RESPACK.
Some of the calculations were performed by using
the Supercomputer Center, located at the Institute for Solid State Physics, the University of Tokyo.
A part of this work is financially supported by TOYOTA MOTOR CORPORATION.
The authors were supported by Building of Consortia for the Development of 
Human Resources in Science and Technology from the MEXT of Japan. 
This work has been also supported by a Grant-in-Aid for Scientific Research 
(Nos. 16H06345 and 20H01850) from Ministry of Education, Culture, Sports, Science and Technology, Japan.  

\clearpage
\appendix
\section{Features of Descriptors}\label{App}
Here, we summarize the features of the descriptors
used in constructing the regression models and the PCA.
In constructing the regression models,
we prepare 72 models based on the seven choices shown in Table \ref{reg_desc_table}.

\begin{table*}[t]
\caption{Choices of descriptors used in the regression model. Our choices when constructing the best regression model are shown in the second column.}
\begin{tabular}{lc}
\toprule
  Choices of Descriptors  & Best model                     \\
\midrule
  Normalizing all the descriptors except for $t_{\rm max, all}$ by $t_{\rm max, all}$ & True\\
  Using square of descriptors (e.g., $U_{\rm mean, diag}^2$) & True \\
  Using cross terms (e.g., $U_{\rm mean, diag}\times t_{\rm max, all}$) & True \\
  Including $V_{\rm max,diag},V_{\rm min,diag},V_{\rm mean,diag}$ & False \\
  Including $J_{\rm max,offdiag},J_{\rm min,offdiag},J_{\rm mean,offdiag}$ & False \\
  Including $V_{\rm mean, diag}/U_{\rm max, diag}$ when $V_{\rm mean,diag}$ is used as the descriptor & -- \\
  Including $J_{\rm mean, offdiag}/U_{\rm max, diag}$ when $J_{\rm mean,offdiag}$ is used as the descriptor & -- \\
\bottomrule
\end{tabular}
\label{reg_desc_table}
\end{table*}

In addition, all the models include $t_{\rm p, all}$(p=mean,max) and $U_{\rm p, diag}$ (p=mean, max, min).
We standardized these descriptors for the construction of the regression model.
In Table~\ref{desc_table},
we list the descriptors used in PCA
in descending order of the absolute value corresponding to the components of the first principal vector, $|v^1_i|$.

\begin{table}[t]
\caption{Descriptors used in PCA}
\begin{tabular}{ll}
\toprule
  {$i$}  &~~~~~descriptor                     \\
\midrule
  0 &~~~~~$   V_{\rm min, diag} $             \\
  1 &~~~~~$   V_{\rm min, offdiag} $          \\
  2 &~~~~~$   V_{\rm mean, diag}  $           \\
  3 &~~~~~$   V_{\rm mean, all}   $           \\
  4 &~~~~~$   V_{\rm mean,offdiag}  $         \\
  5 &~~~~~$   V_{\rm max,offdiag}  $          \\
  6 &~~~~~$   V_{\rm max,diag}     $          \\
  7 &~~~~~$   V^{\prime}_{\rm min,diag} $     \\
  8 &~~~~~$   V^{\prime}_{\rm min,offdiag}$   \\
  9 &~~~~~$   V^{\prime}_{\rm mean,diag}  $   \\
  10 &~~~~~$   V^{\prime}_{\rm mean,all}   $  \\
  11 &~~~~~$   V^{\prime}_{\rm mean,offdiag}$ \\
  12 &~~~~~$   V^{\prime}_{\rm max,offdiag} $ \\
  13 &~~~~~$   V^{\prime}_{\rm max,diag}    $ \\
  14 &~~~~~$   U_{\rm min,offdiag}    $       \\
  15 &~~~~~$   U_{\rm mean,offdiag} $         \\
  16 &~~~~~$   U_{\rm max,offdiag}  $         \\
  17 &~~~~~$   U_{\rm mean,all}     $         \\
  18 &~~~~~$   U_{\rm mean,diag}    $         \\
  19 &~~~~~$   U_{\rm max,diag}     $         \\
  20 &~~~~~$   U_{\rm min,diag}     $         \\
\midrule
  21 &~~~~~$   J_{\rm mean,offdiag} $        \\
  22 &~~~~~$   J_{\rm max,offdiag}  $        \\
  23 &~~~~~$   t_{\rm mean,offdiag} $         \\
  24 &~~~~~$   t_{\rm max,diag}     $         \\
  25 &~~~~~$   J_{\rm min,offdiag}  $         \\
  26 &~~~~~$   t_{\rm min,diag}     $         \\
  27 &~~~~~$   t_{\rm mean,all}     $         \\
  28 &~~~~~$   t^{\prime}_{\rm min,diag}    $ \\
  29 &~~~~~$   t^{\prime}_{\rm max,offdiag} $ \\
  30 &~~~~~$   t^{\prime}_{\rm mean,offdiag}$ \\
  31 &~~~~~$   t_{\rm mean,diag}         $    \\
  32 &~~~~~$   t_{\rm max,offdiag}       $    \\
  33 &~~~~~$   t^{\prime}_{\rm max,diag} $    \\
  34 &~~~~~$   t^{\prime}_{\rm mean,all} $    \\
  35 &~~~~~$   t^{\prime}_{\rm mean,diag}$    \\
\bottomrule
\end{tabular}
\label{desc_table}
\end{table}

\section{Construction of regression model with undersampling}\label{App2}

  In this appendix, we discuss the effect of undersampling 
on our database. 
Our database listed in Table \ref{tc_table} contains 29 materials, but 
almost half of them are labeled as 1111 compounds. 
Such an imbalance in the dataset sometime causes 
overfitting problems 
in constructing the regression model.

\begin{figure}[bt!]
  \begin{center}
    \includegraphics[width=8cm,clip]{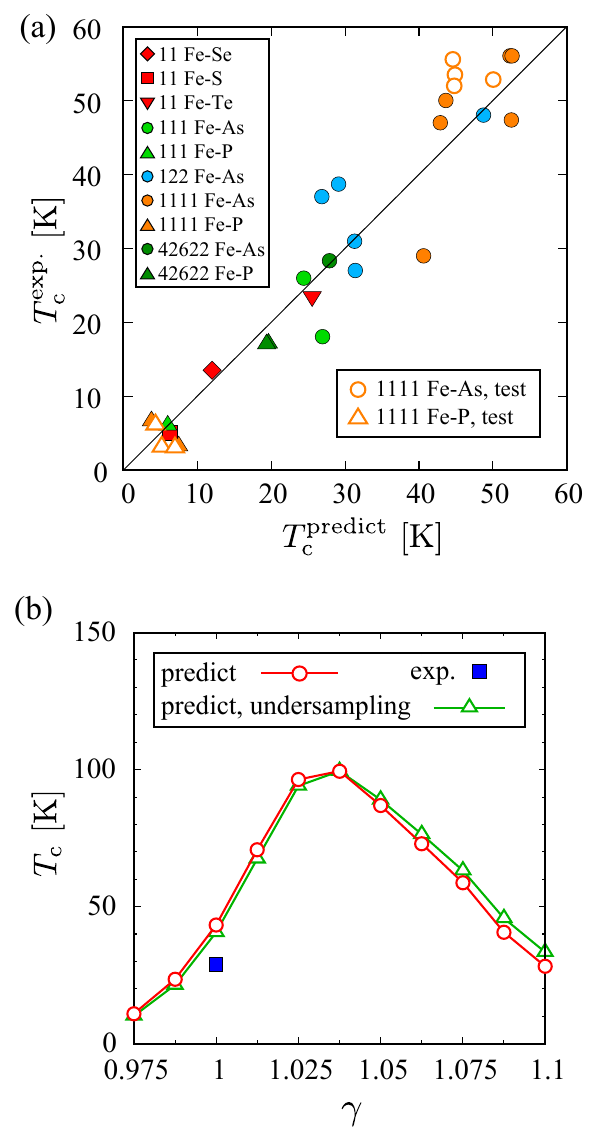}
  \end{center}
  \caption{
  (color online)
  (a) Experimental $T_\text{c}$ vs. predicted $T_\text{c}$ obtained from the 
  regression model with the undersampling, which removes 
  four 1111-As compounds (GdFeAsO, NdFeAsO, SmFeAsO, PrFeAsO) and three 1111-P compounds (GdFePO, NdFePO, SmFePO) from the training data. 
  Open symbols mean the results for 1111 compounds treated as the test data. 
  The other notations are the same as those of Fig.~\ref{reg}.
  The coefficient determination $R^2$ is about 0.90.
  (b) $T_\text{c}$ for LaFeAsO with hypothetical structures predicted by the regression model.
  Red circles and blue squares are the same results plotted in Fig. \ref{reg}.
  Green triangles represent the \Tc of LaFeAsO with hypothetical structures predicted by using the model with the undersampling.
  }
\label{undersampling}
\end{figure}

  To check whether the overfitting happens or not, we construct the regression model 
using the database with undersampling, 
which means that we remove several 1111 compounds from the training data.
Figure \ref{undersampling}(a) shows the accuracy of our regression model with the undersampling.
We find that the model with the undersampling also reproduces the experimental \Tc data with 
$R^2 \approx 0.90$.
The predicted \Tc for 1111 compounds treated as the test data well reproduce the experimental ones.
This model also predicts almost the same results for \Tc for LaFeAsO with a hypothetical structure, 
shown in Fig. \ref{undersampling} (b).
This result suggests that the overfitting does not occur in the predicted results 
in the main text, and thus, the original dataset without the undersampling 
is reasonable.

\bibliography{reference}

\begin{thebibliography}{100}
\bibitem{note1}
$T_\text{c}^\text{exp.}$ is estimated from a figure where the temperature
  dependence of the electronic specific heat is shown.
\bibitem{note2}
${\rm LaFeAsO_{1-x}H_{x}}$ has the two-dome $T_\text{c}$ structure.
  $T_\text{c}^\text{exp.}$ is used the maximum $T_\text{c}$ of the dome in the
  low-doping region.
\bibitem{note3}
It is not fully clarified whether the system shows the bulk superconductivity.
\end{thebibliography}

\end{document}